\documentclass[
    ,final            % use final for the camera ready runs
  ]
  {aipproc}

\layoutstyle{6x9}

\newcommand{\kms}{km~s$^{-1}$ }

\begin{document}

\title{RV survey of low-mass companions to sdB stars}


\author{Lew Classen}{
  address={Dr.~Karl Remeis--Observatory \& ECAP, Astronomical Institute,
Friedrich-Alexander University Erlangen-Nuremberg, Sternwartstr. 7, D-96049 Bamberg, Germany}
}

\author{Stephan Geier}{
  address={Dr.~Karl Remeis--Observatory \& ECAP, Astronomical Institute,
Friedrich-Alexander University Erlangen-Nuremberg, Sternwartstr. 7, D-96049 Bamberg, Germany}
}

\author{Ulrich Heber}{
  address={Dr.~Karl Remeis--Observatory \& ECAP, Astronomical Institute,
Friedrich-Alexander University Erlangen-Nuremberg, Sternwartstr. 7, D-96049 Bamberg, Germany}
}

\author{Simon J. O'Toole}{
  address={Australian Astronomical Observatory, PO Box 296, Epping, NSW, 1710, Australia}
}

\begin{abstract}
 The origin of subdwarf B (sdB) stars is not fully understood yet since it requires high mass loss at the red giant stage. SdBs in close binary systems are formed via common envelope ejection, but the origin of apparently single sdB stars remains unclear. Substellar companions may be able to trigger common envelope ejection and help forming sdBs that appear to be single. \\
Using a sample of high resolution spectra we aim at detecting small radial velocity (RV) shifts caused by such low mass (sub-)stellar companions. The RVs are measured with high accuracy using sharp metal lines. Our goal is to test the theoretical predictions and put constraints on the population of the lowest mass companions to sdB stars.
\end{abstract}

\maketitle

%%%%%%%%%%%%%%%%%%%%%%%%%%%%%%%%%%%%%%%%%%%%
%% MAINMATTER
%%%%%%%%%%%%%%%%%%%%%%%%%%%%%%%%%%%%%%%%%%%%

\section{Introduction}

Subluminous stars are considered to be core helium-burning objects with a thin hydrogen shell. In a Hertzsprung-Russell diagram these stars are
situated on the Extreme Horizontal Branch (EHB). The evolution of these stars is far from understood since the formation of an sdB requires
extreme mass loss at the tip of the RGB. The mechanism leading to this mass loss is still under debate \citep{Heber1986, Heber2009}.

Formation scenarios can be roughly divided into single star and binary channels. About 50\% of the known sdBs are found to be in close binaries with periods of few hours to several days consistent with the predictions of the common envelope (CE) ejection scenario \citep{Han2002}.
Two main sequence stars evolve in a binary system. The more massive one becomes a red giant first and fills its Roche Lobe. Unstable mass transfer leads to the formation of a CE. The core of the red giant and the immersed secondary experience a loss of orbital energy which is deposited in the CE causing a spiral-in towards the center of mass. As soon as enough energy has been deposited in the envelope, it is ejected. The inwards migration stops and the stars end up in a close binary. This scenario is well suited to explain the known population of close sdB binaries.

But how are the apparently single sdBs formed? \citet{Soker1998} suggested that substellar companions like planets or brown dwarfs may be able to trigger CE ejection as well. In order to survive the common envelope and avoid evaporation inside the envelope the secondary should have a minimum mass of more than $10\,M_{\rm J}$.

A similar scenario has been proposed for the formation undermassive single white dwarfs by \citet{Nelemans1998}.

Several substellar objects have recently been detected around hot subdwarfs in wide orbits, showing that planets can survive the red giant phase of their host star \citep{Silvotti2007, Lee2009, Qian2009}. Recently, \citet{Geier2009} discovered a candidate substellar companion in close orbit around the bright sdB HD~149382, which perfectly fits into the pattern predicted by Soker. Systems like that may be quite common and could have remained unnoticed up to now.

The goal of our project is to put constraints on the population of the lowest mass companions to sdB stars by measuring accurate radial velocities from multi-epoch high resolution spectra.

\begin{figure}[!t]
  \includegraphics[height=0.95\textwidth,angle=-90]{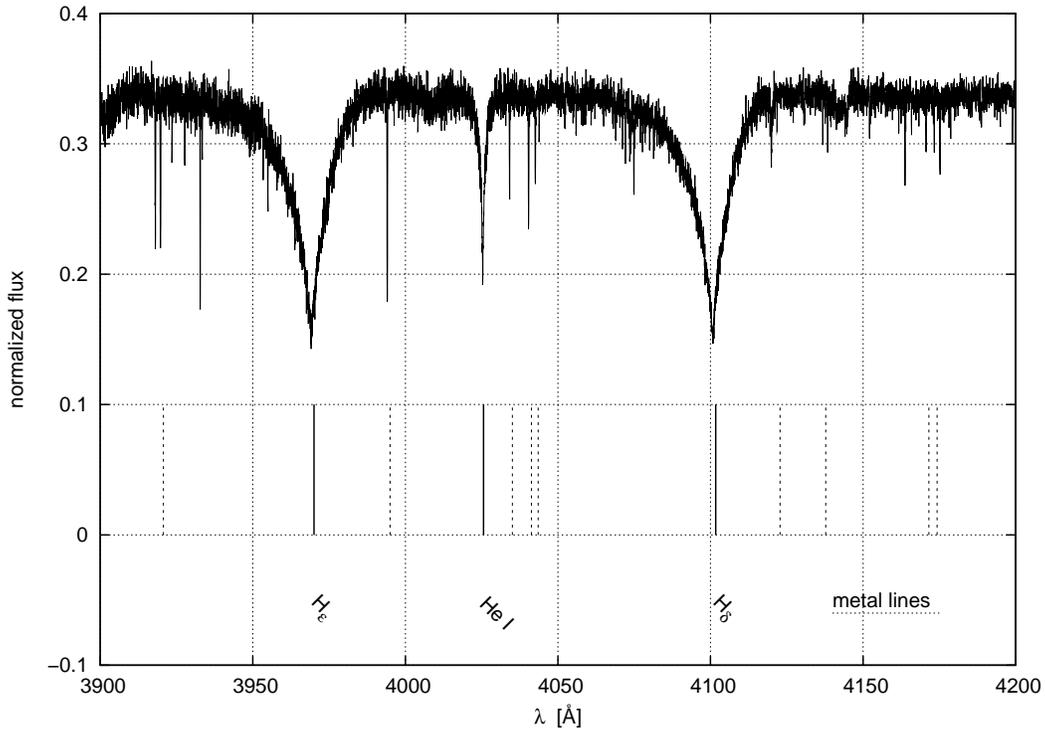}
  \caption{The plot displays a section of a FEROS spectrum of HD~205805 featuring exemplary sharp metal absorption lines.}
\label{fig:uebers}
\end{figure}

\section{Observations and data reduction}

Up to now our survey consists of 23 bright sdB stars with visual magnitudes ranging from $10$ to $14$. We used high resolution spectra obtained with the FEROS instrument mounted at the ESO/MPG-2.2m telescope. The spectra have a resolution of $48\,000$. Spectra were reduced, calibrated and corrected for earth's orbital motion with the MIDAS package using the FEROS reduction pipeline.

All programme stars are single-lined objects without visible spectral features of a companion. Furthermore, those binaries with high radial velocity variability have been excluded because their unseen companions  must be quite massive. Each star has been observed several times (from 2 to 15) on timescales ranging from one day to some years. As an example we display a section of the FEROS spectrum of HD~205805 in Fig \ref{fig:uebers}.

\begin{figure}[!t]
  \includegraphics[scale=0.6,angle=-90]{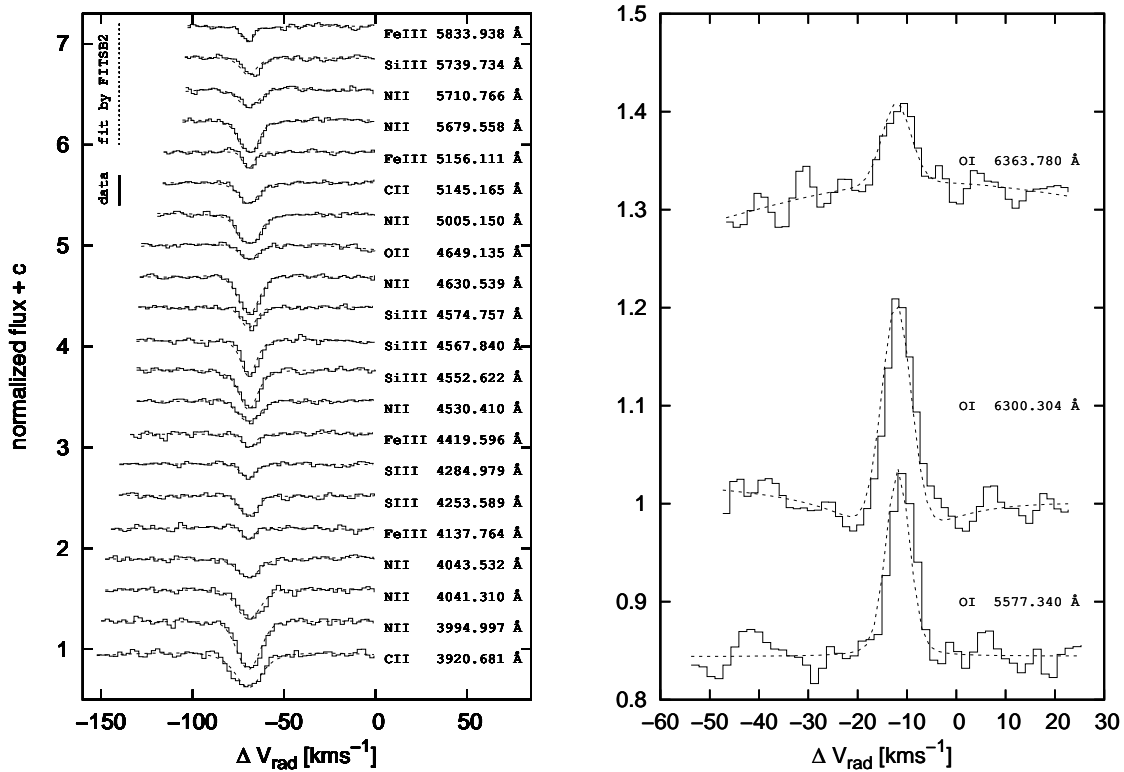}
  \caption{Spectral features of HD~205805 fitted with FITSB2.
The left hand panel displays metal absorption lines suitable for RV measurements, in the right hand panel exemplary night-sky emission lines  used to check the wavelength calibration are shown. The latter are shifted in RV due to heliocentric correction applied to the spectra.}
\label{fig:fitlines}
\end{figure}

\section{Measurement}

Subdwarf B stars are hot ($T_{\rm eff}=20\,000-35\,000\,{\rm K}$) and their atmospheres radiative. The do\-mi\-nant features in their spectra are the hydrogen Balmer and several helium lines. But the spectra also show up to $\simeq50$ weak metal absorption lines which are very sharp. These features are best suited to measure radial velocities.

The atmospheres of sdB stars are considered to be stable. The only known effect other than a close  companion that may lead to RV variability are pulsations. Pulsating sdB stars do indeed exist with periods between $90\,{\rm s}$ and $3\,{\rm h}$ and RV amplitudes of few \kms at most \citep{Heber2009}.

For our analysis we chose a set of sharp, unblended metal lines situated between $3600$ and $6600$\,\AA  \,(see Fig. \ref{fig:fitlines}, left hand panel). Accurate rest wavelengths were taken from the NIST database. Depending on the atmospheric parameters of the star and the quality of the data a subset was used.

The RV measurement consisted of two steps. First, we fitted Gaussian and Lorentzian profiles to each line separately using the SPAS routine (Hirsch priv. comm.). The resulting mean RVs had standard deviations ranging from $0.8$ to $2.7\,{\rm km\,s^{-1}}$. Second, we used the FITSB2 routine \citep{Napiwotzki2004} to perform a simultaneous fit of all suitable lines. Uncertainties were calculated using a bootstrapping algorithm and found to range from $0.1$ to $0.5\,{\rm km\,s^{-1}}$, while both methods result in consistent mean RVs.

The scatter of individual RV values measured with SPAS around the mean value was reviewed for possible systematic effects like correlations with certain elements. No such effects could be found.
To check the wavelength calibration for systematic errors we used telluric features as well as night-sky emission lines (see Fig. \ref{fig:fitlines}, right hand panel). Having their origin on earth these lines should have zero RV. The FEROS instrument turned out to be very stable. Usually corrections of less than $0.5\,{\rm km\,s^{-1}}$ had to be applied.

In order to derive the degree of precision we can achieve with our method, we generated synthetic model spectra with realistic atmospheric parameters including Poisson noise with $S/N\,=\,1000$ and determined the RV in the way described above. The uncertainties turned out to range from $0.1$ to $0.4\,{\rm km\,s^{-1}}$. The accuracy is limited by the resolution of the spectrograph, the number of features and the intrinsic thermal broadening of the lines.

\section{Results}

\begin{figure}[!h]
  \includegraphics[height=\textwidth,angle=-90]{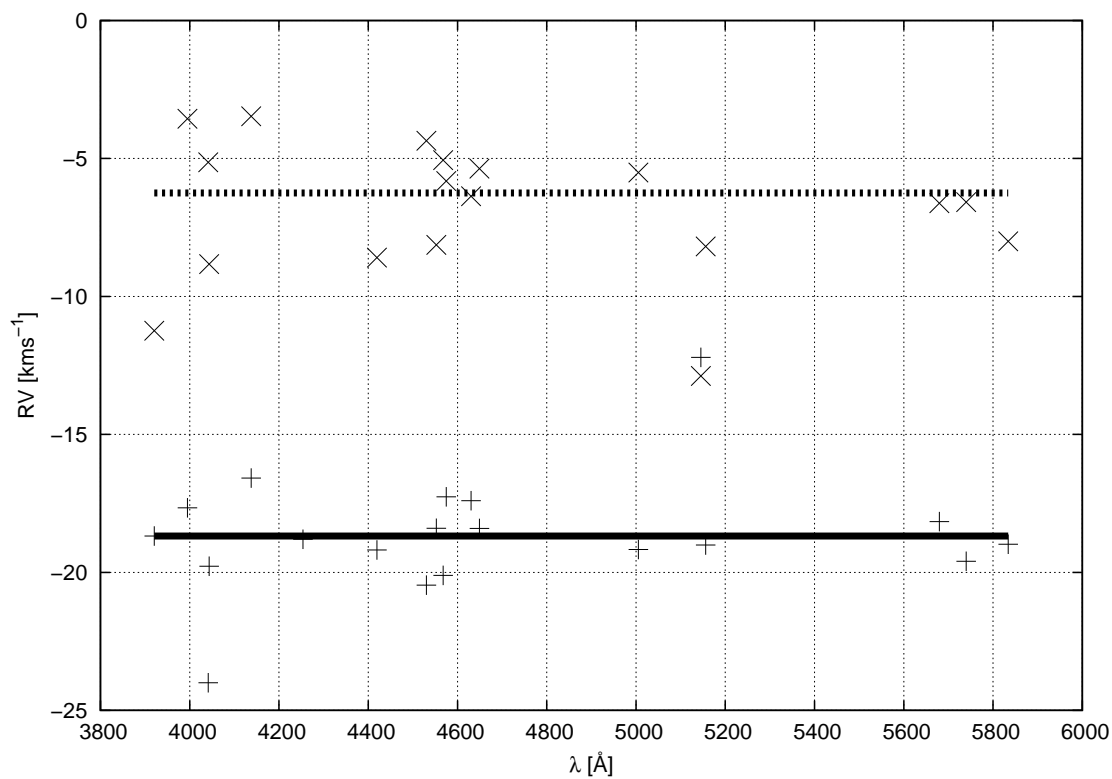}
  \caption{The diagram shows the result of RV determination for two spectra of EC~20106$-$5248 taken 6 months apart. Each measuring point represents a metal absorption line. The horizontal lines marking mean RV values of the spectra indicate a significant shift in radial velocity.}

\label{fig:bigshift}
\end{figure}

\pagebreak

Four stars were found to be significantly (another two marginally) RV variable. 
The shift in RV ranges from $0.5$ to $12.7\,{\rm km\,s^{-1}}$ (see Tab. \ref{tab:a}). As an example we display two RV measurements for EC~20106$-$5248 in Fig. \ref{fig:bigshift}. Follow-up photometry  and high resolution spectroscopy are necessary to exclude pulsational variability and derive the orbital parameters of these binaries. Constraints can then be put on the companion masses. Even if there are no low mass companions we would expect to find a certain fraction of systems with small RV shifts caused by more massive secondaries seen at low inclination. A larger sample is needed to exclude this case.

\begin{table}[t]
\begin{tabular}{lrrrr}
\hline
 & SPAS &  & FITSB2 & \\
\hline
    \tablehead{1}{l}{b}{star}
  & \tablehead{1}{r}{b}{ $\Delta \mathbf{V_{\rm rad\, max}}$ [\kms] }
  & \tablehead{1}{r}{b}{ standard \\deviation}
  & \tablehead{1}{r}{b}{$\Delta \mathbf{V_{\rm rad\, max}}$ [\kms] }
  & \tablehead{1}{r}{b}{mean error}   \\
\hline
synthetic & $0.06$ & $0.28$	& $0.10$ & $0.00$\\
EC 00042$-$2737 & $0.08$ & $1.21$ & & \\
CD-3515910 & $0.17$ & $1.11$ & & \\
Feige 38 & $0.21$ & $0.93$ & & \\
EC 01120$-$5259 & $0.23$ & $1.23$ & & \\
PG 1432$+$004 & $0.32$ & $1.11$ & & \\
EC 02542$-$3019 & $0.35$ & $2.68$ & & \\
PG 1505$+$074 & $0.40$ & $0.64$ & & \\
EC 20229$-$3716 & $0.41$ & $0.79$ & & \\
EC 21043$-$4017 & $0.50$ & $1.30$ & & \\
EC 03591$-$3232 & $0.53$ & $1.19$ & & \\
HD 4539 & $0.64$ & $0.91$ & & \\
HD 205805 & $0.71$ &  $1.00$ & $0.62$ & $0.09$\\
PHL 334 & $0.84$ & $1.08$ & & \\
EC 11349$-$2753 & $1.04$ & $0.95$ & $1.02$ & $0.13$\\
PG 0342$+$026 & $1.06$ & $0.94$ & & \\
EC 14345$-$1729 & $1.10$ & $1.30$ & & \\
HE 0151$-$3919 & $1.29$ & $1.65$ & & \\
PG 2151$+$100 & $2.08$ & $1.36$ & & \\
EC 15103$-$1557 & $2.60$ & $1.58$ & $0.80$ & $0.52$\\
PHL 44 & $3.16$ & $1.76$ & & \\
JL 36 & $5.38$ & $1.96$ & $3.20$ & $0.49$\\
EC 13506$-$3137 & $5.64$ & $1.52$ & & \\
EC 20106$-$5248 & $12.43$ & $1.61$ & $12.65$ & $0.22$\\
\hline
\end{tabular}
\caption{Summary of RV results for our sample obtained using two different algorithms. RVs for each line were measured independently and averaged with SPAS (left hand column) whereas FITSB2 uses a simultaneous fit (right hand column).}
\label{tab:a}
\end{table}

\begin{figure}[t]
  \includegraphics[height=0.95\textwidth,angle=-90]{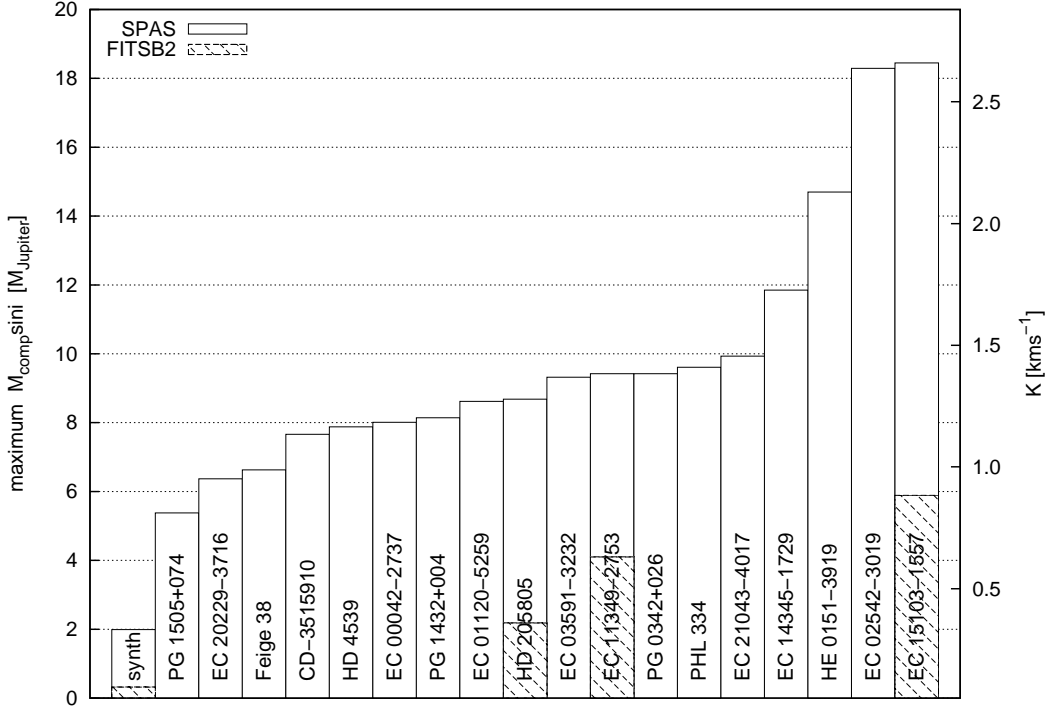}
  \caption{The histogram shows upper mass limits for unseen companions in close orbits around sdBs, where no significant RV shifts could be detected (left hand axis). These are derived from upper limits for the RV semi-amplitudes (right hand axis).}
\label{fig:const}
\end{figure}

No RV variations were found in the rest of the sample. Nevertheless, the errors allow to derive upper limits for $M_{\rm comp}\,{\rm sin}\,i$ of undetected substellar companions in close orbits from the binary mass function (Eq. \ref{eq:1}). 

\begin{equation} \label{eq:1}
f(M)=\frac{M^{3}_{\rm comp}\sin^3 i}{(M_{\rm comp}+M_{\rm sdB})^2}=\frac{PK^3}{2 \pi G}
\end{equation}

\citet{Soker1998} suggested that these objects should be more massive than $10\,M_{\rm J}$ and should have orbital periods shorter than $10\,{\rm d}$. Adopting this maximum period and assuming a canonical mass of $0.47\,M_{\rm \odot}$ for the sdB, the errors of the RV measurements can be used as upper limits for the RV semi-amplitudes ($K$).

As can be seen in Fig. \ref{fig:const}, tight constraints can be put on possible substellar companions, if the measurement error is smaller than $\simeq1.0\,{\rm km\,s^{-1}}$. Given data of perfect quality, close-in planets with masses as low as $0.3\,M_{\rm J}$ can be firmly excluded if our assumption about the orbital period is correct.

In conclusion $\simeq\,26\%$ of the apparently single stars in our sample show small RV variations. The true fraction is expected to be higher, because the accuracy is limited by the S/N in most cases. Companions with masses higher than $10\,M_{\rm J}$ in close orbits ($P\,<\,10\,{\rm d} $) can be excluded for $61\%$ of our sample.

One possible explanation for the fractions found is that the assumed CE scenario is not the only channel for sdB formation. However, even with CE ejection being the only channel, the companion may be destroyed during the CE phase either by evaporation or a merger with the stellar core leaving behind a single sdB. 
Better statistics and reliable predictions from theoretical calculations are required to solve this problem. 
%Nevertheless our work demonstrates the suitability of sdBs for accurate RV measurements showing that reliable results can be obtained by %quite simple methods. 

%%%%%%%%%%%%%%%%%%%%%%%%%%%%%%%%%%%%%%%%%%%%%%%%
%% BACKMATTER
%%%%%%%%%%%%%%%%%%%%%%%%%%%%%%%%%%%%%%%%%%%%%%%%

%\pagebreak

\begin{theacknowledgments}
 S.G. is supported by DFG through grant He 1353/49-1.
\end{theacknowledgments}

%%%%%%%%%%%%%%%%%%%%%%%%%%%%%%%%%%%%%%%%%%%%%%%%
%% The bibliography can be prepared using the BibTeX program or
%% manually.
%%
%% The code below assumes that BibTeX is used.  If the bibliography is
%% produced without BibTeX comment out the following lines and see the
%% aipguide.pdf for further information.
%%
%% For your convenience a manually coded example is appended
%% after the \end{document}
%%%%%%%%%%%%%%%%%%%%%%%%%%%%%%%%%%%%%%%%%%%%%%%%

%%%%%%%%%%%%%%%%%%%%%%%%%%%%%%%%%%%%%%%%%%%%%%%%
%% You may have to change the BibTeX style below, depending on your
%% setup or preferences.
%%
%%
%% For The AIP proceedings layouts use either
%%%%%%%%%%%%%%%%%%%%%%%%%%%%%%%%%%%%%%%%%%%%

\newpage

\bibliographystyle{aipproc}   % if natbib is available
%\bibliographystyle{aipprocl} % if natbib is missing

%%%%%%%%%%%%%%%%%%%%%%%%%%%%%%%%%%%%%%%%%%%
%% You probably want to use your own bibtex database here
%%%%%%%%%%%%%%%%%%%%%%%%%%%%%%%%%%%%%%%%%%%
%\bibliography{litverzeichnis}

\end{document}